\documentclass{iopart} 
\usepackage{graphics}
\usepackage{epsfig}
\newcommand{\be}[1]{\begin{equation}\label{#1}}
\newcommand{\ee}{\end{equation}}   
\newcommand{\fig}[1]{figure~\ref{fig#1}}  
\newcommand{\bea}{\begin{eqnarray}}
\newcommand{\eea}{\end{eqnarray}} 
\newcommand{\eq}[1]{(\ref{#1})}

\begin{document}  
\jl 2
\title{\bf Inelastic semiclassical Coulomb scattering}
\author{Gerd van de Sand\dag\ and  Jan M Rost\ddag}
\address{\dag-- Theoretical Quantum Dynamics --\\
Fakult\"at f\"ur Physik, Universit\"at Freiburg,
Hermann--Herder--Str.  3, D--79104 Freiburg,
Germany}
\address{\ddag Max-Planck-Institute for Physics of Complex Systems, 
N\"othnitzer Str. 38, D-01187 Dresden,
Germany}
\date{\today}
\begin{abstract}
We present a semiclassical S-matrix study of  inelastic collinear
electron-hydrogen scattering. A simple way to extract all necessary
information from the deflection function alone without having to
compute the stability matrix is described. This includes the determination
of the relevant Maslov indices. Results of singlet and triplet cross sections
for excitation and ionization are reported. The different levels
of approximation -- classical, semiclassical, and uniform semiclassical --
are compared among each other and to the full quantum result.
\end{abstract}

\section{Introduction}

Semiclassical scattering theory was formulated almost  40 years ago for
potential scattering in terms of WKB-phaseshifts \cite{FoW59}.
Ten years later, a multidimensional  formulation appeared, derived from the 
Feynman path integral  \cite{Pec69}.  Based on a similar derivation
Miller developed at about the same time his 'classical S-matrix' which
extended  Pechukas' multidimensional semiclassical S-matrix  for potential
scattering to inelastic scattering \cite{Mil70,Mil70a,Mil74}.
These semiclassical concepts have been mostly applied to molecular 
problems, and in a parallel development by Balian and Bloch \cite{BaBl74}
to condensed matter problems, i.e. to short range interactions.

Only recently, scattering involving long range (Coulomb) forces has been
studied using semiclassical S-matrix techniques, in particular 
potential scattering \cite{RoHe94}, ionization of atoms near the threshold
\cite{Ros94,Ros95} and chaotic scattering below the ionization threshold
\cite{RoWi96}. The latter problem has also been studied purely classically
\cite{GuY93} and semiclassically within a periodic orbit approach
\cite{ERTW91}.

While  there is a substantial body of work on classical collisions
with Coulomb forces using the Classical Trajectory Monte Carlo Method (CTMC) 
almost no semiclassical studies exist. This fact together with the remarkable
success of CTMC methods have motivated our semiclassical investigation
of inelastic Coulomb scattering.  To carry out an explorative study 
 in the full (12) dimensional
phase space of three interacting particles is prohibitively 
expensive. Instead, we restrict ourselves
to {\it collinear} scattering, i.e. all three particles  are located on a line
with the nucleus in between the two electrons. This collision configuration
was proven to contain the essential physics for ionization near the threshold
\cite{Ros94,Wan53,Ros97} and it fits well into the context of classical 
mechanics since the collinear phase space is the consequence of a stable
partial fixed point at the interelectronic angle $\theta_{12} = 180^\circ$
\cite{Ros97}. Moreover, it is exactly the setting of Miller's approach 
for molecular reactive scattering. 

For the theoretical development of scattering concepts another  Hamiltonian
of only two degrees of freedom has been established in the literature,
the s-wave model \cite{HDIF93}. Formally, this model Hamiltonian is obtained
by averaging the angular degrees    of freedom and retaining only the 
zeroth order of the respective multipole expansions. The resulting
electron-electron interaction is limited to the line $r_1 = r_2$, where
the $r_i$ are the electron-nucleus distances, and   the potential is
not differentiable along the line $r_1 = r_2$. This is not very
important for the quantum mechanical treatment, however, it affects the
classical mechanics drastically. Indeed, it has been found that
the s-wave Hamiltonian leads to a threshold law for ionization 
very different from the one resulting 
from the collinear and the full Hamiltonian (which both lead to the same
threshold law) \cite{FIM99}. 
Since it is desirable for a comparison of semiclassical with quantum results
that the underlying classical mechanics does not lead to qualitative different
physics we have chosen to work with the collinear Hamiltonian.
For this collisional system we will obtain and compare the classical, 
the quantum and the primitive and uniformized semiclassical result.
For the semiclassical calculations the collinear Hamiltonian was 
amended by the so called Langer correction, introduced by Langer 
\cite{Lan37} to overcome inconsistencies with the semiclassical 
quantization  in spherical (or more generally non-cartesian)
coordinates.

As a side product of this study we give a  rule how to obtain 
the correct Maslov indices for a two-dimensional collision system 
directly from the deflection function without the stability matrix. This does not only make the semiclassical 
calculation much more transparent it also  considerably reduces the numerical 
effort since one can avoid to compute the stability matrix and 
nevertheless one obtains the full semiclassical result. 

The plan of the paper is as follows: in section 2 we introduce the Hamiltonian
and the basic semiclassical formulation of the S-matrix in terms of classical
trajectories. We will discuss a 
typical S-matrix $S(E)$ at fixed total energy $E$ and illustrate a simple way to 
determine the relevant (relative) Maslov phases.
In section 3 semiclassical excitation and ionization probabilities
are compared to quantum results for singlet and triplet symmetry.
The spin averaged probabilities are also compared to the classical results. 
In section 4 we will go one step further and uniformize the semiclassical
S-matrix,  the corresponding scattering probabilities will be presented.
We conclude the paper with section 5  where we try to assess how useful
semiclassical scattering theory is for Coulomb potentials. 
\section{Collinear electron-atom scattering}
\subsection{The Hamiltonian and the scattering probability}
 The collinear two-electron Hamiltonian with a proton as a nucleus reads
 (atomic units are used throughout the paper)
 \be{ham}
 h = \frac{p_{1}^{2}}{2}+ \frac{p_{2}^{2}}{2} -\frac{1}{r_{1}}
 -\frac{1}{r_{2}} -\frac{1}{r_{1}+r_{2}}.
 \ee
The Langer-corrected Hamiltonian reads  
\be{hlanger}
H = h + \frac{1}{8r_{1}^{2}}+ \frac{1}{8r_{2}^{2}}.
\ee
For collinear collisions  we have only one 'observable' after the collision,
namely the state with quantum number $n$, 
to which the target electron was excited through the
collision. If its initial quantum number before the collision was $n'$,
we may write the probability at  total energy $E$ as
\be{prob}
P_{n,n'}(E) = |\langle n|S|n'\rangle|^{2}
\ee
with the S-matrix
\be{S-matrix}
S = \lim_{t \to \, \infty \atop t' \to -\infty}
e^{iH_{f}t}e^{-iH(t-t')}e^{-iH_{i}t'}.
\ee
Generally, we use the prime to distinguish initial from final state
variables. The Hamiltonians $H_{i}$ and $H_{f}$  represent the scattering
system before and after the interaction and do not need to be identical
(e.g. in the case of a rearrangement collision). The initial energy
of the projectile electron is given by 
\be{encon}
\epsilon' =  E - \tilde\epsilon'
\ee
where $\tilde\epsilon'$ is the energy of the bound electron and
$E$ the total energy of the system. In the same way the final energy of the free
electron is  fixed. However, apart from excitation,  ionization can also occur
for $E>0$  in which case $|n\rangle$ is simply replaced by by a free momentum
state $|p\rangle$. This is possible since the complicated asymptotics of three
free charged particles in the continuum is contained in the S-matrix.

\subsection{The semiclassical expression for the S-matrix}
Semiclassically, the S-matrix  may be expressed as 
\be{S-m}
S_{n,n'}(E) = \sum_{j}\sqrt{{\cal P}_{n,n'}^{(j)}(E)} \, 
e^{i\Phi_{j}-i\frac{\pi}{2}\nu_{j}},
\ee
where the sum is over all classical trajectories $j$ which connect the
initial state $n'$ and the final 'state' $n$ with  a respective probability of
${\cal P}_{n,n'}^{(j)}(E)$. The classical probability ${\cal P}_{n,n'}^{(j)}(E)$ is given by
\be{probcl}
{\cal P}_{n,n'}^{(j)}(E) = {\cal P}_{\epsilon,\epsilon'}^{(j)}(E) 
\, \frac{\partial \epsilon}{\partial n} = 
\frac{1}{N} \,  \left| \frac{\partial \epsilon(R')}{\partial R_j'} \right|^{-1} 
\frac{\partial \epsilon}{\partial n}\, ,
\ee
see \cite{Ros95} where also an expression for the normalization constant $N$ is 
given. Note, that due to the relation \eq{encon} derivatives of $\epsilon$
and $\tilde\epsilon$ with respect to $n$ or $R'$ differ only by a sign.
From now on we denote the coordinates of the initially free electron by capital letters
and those of the initially bound electron by small letters. 
If the projectile is bound after the collision we will call this an 'exchange 
process', otherwise we speak of 'excitation' (the initially
bound electron remains bound)
or ionization (both electrons have positive energies).
The deflection function $\epsilon(R')$ has to be calculated numerically, 
as described in the next section.
The phase $\Phi_j$ is the collisional action \cite{Coul} given by
\be{colacang}
\Phi_j \left(P,n;P',n' \right) = -\int \, dt \, \left( q \dot{n} + R \dot{P} \right)
\ee
with the angle variable $q$. The Maslov index $\nu_{j}$ counts the number of caustics
along each trajectory.  
'State' refers in the present  context to  integrable
motion for asymptotic times $t\to\pm\infty$, characterized by  constant
actions, $J' =  2\pi\hbar(n' + 1/2)$. The (free) projectile  represents
trivially integrable motion and can be characterized by 
its momentum $P'$.
In our case,  each particle has only one degree of freedom. Hence, instead
of the action $J'$ we may use the energy $\tilde\epsilon'$ for a unique
determination of the initial bound state.
In the next sections we describe how we calculated the deflection function,
the collisional action and the Maslov index.
\subsubsection{Scattering trajectories and the deflection function}
The crucial object for the determination of (semi-)classical 
scattering probabilities
is the deflection function ${\epsilon}(R')$ where ${\epsilon}$ 
is the final energy of the projectile electron as a function of its initial
position $R_0 + R'$. Each trajectory is started with the bound electron
at an arbitrary but fixed phase space point on the
 degenerate Kepler ellipse with  energy 
$\tilde\epsilon'= -1/2$ a.u.. The  initial position of the projectile electron
is changed according to $R'$, but  always at asymptotic distances  (we take
$R_0 = 1000$ a.u.), and its momentum is fixed by energy conservation
to $P' = [2(E-\tilde\epsilon')]^{1/2}$. 
The trajectories are propagated as a function of time with 
a symplectic integrator \cite{Yos90} 
and ${\epsilon} = {\epsilon}(t\to\infty)$  
is in practice evaluated at a time $t$ when
\be{accu}
 d\ln|{\epsilon}|/dt < \delta   
 \ee
   where
$\delta$ determines the desired accuracy of the result. 
Typical trajectories are shown in \fig{1}, their initial
conditions are marked in the deflection function of \fig{2}.

\begin{figure}\hfill
\psfig{file=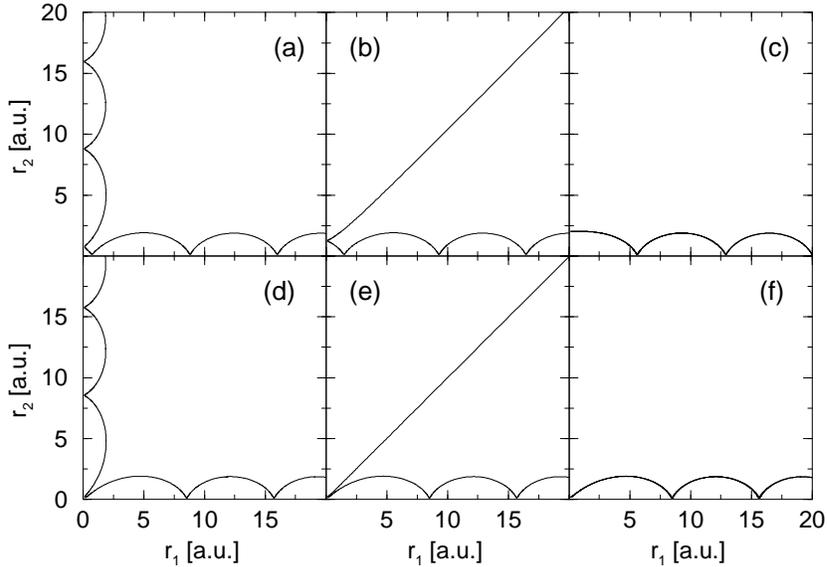,width=11.0cm}
\caption[]{Scattering trajectories at a total energy of
$E=0.125$ a.u.\ with initial conditions marked in 
\protect{\fig{2}}. The labels (a-f) refer to representative 
trajectories with initial values $R'$ shown in \protect{\fig{2}}. The left column
corresponds to classical exchange $n' = 1 \to n = 1$, 
the middle column represents ionization events and the
right column shows elastic back-scattering with 
$n' = 1 \to n = 1$.}
\label{fig1}
\end{figure}

In the present (and generic) 
case of a  two-body potential that is bounded from below 
the deflection function must have maxima and minima 
according to the largest
and smallest energy exchange possible limited by the minimum of the two-body
potential. The deflection function can only be monotonic if the two-body potential is unbounded from below
as in the case of the  pure (homogeneous) Coulomb potential without Langer
correction (compare,  e.g., figure 1 of \cite{Ros94}).
This qualitative difference implies another important
consequence:
For higher total energies $E$ the deflection function is pushed upwards. 
Although energetically allowed, for $ E > 1$ a.u.\ the 
exchange-branch 
vanishes as can be seen from \fig{3}. As we will see later 
this has a significant effect on semiclassical 
excitation and ionization probabilities.

\begin{figure}\hfill
\psfig{file=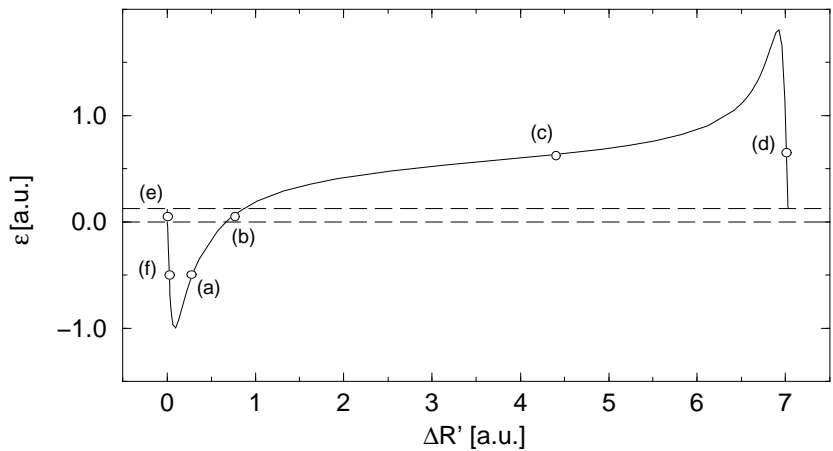,width=11.0cm}
\caption[]{The deflection function at an energy of $E = 0.125$
a.u.\ and for an initial state as described
in the text. The energy interval enclosed by dashed lines marks
ionizing initial conditions and separates the exchange region
(${\epsilon} < 0$) from the excitation region (${\epsilon} > E$),
where ${\epsilon}$ is the energy of the projectile after the collision.}
\label{fig2}
\end{figure}


\begin{figure}\hfill
\psfig{file=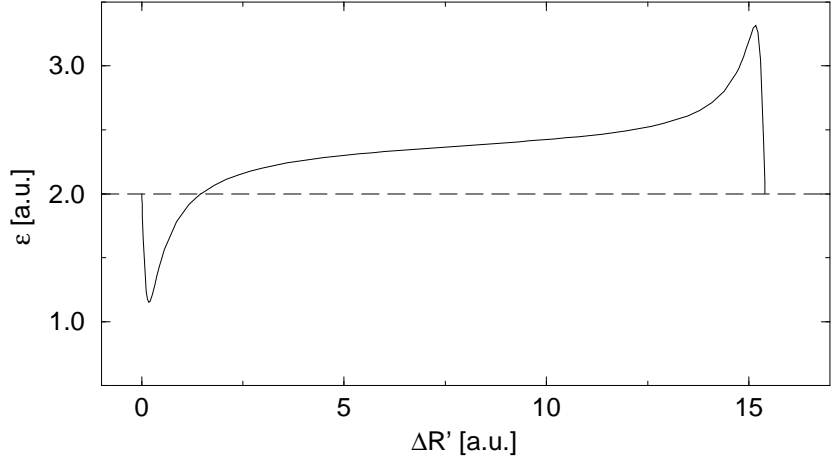,width=11.0cm}
\caption[]{The deflection function at an energy of $E =2$
a.u.\ and for an initial state as described
in the text. The  dashed line separates 
ionizing initial conditions from excitation events.}
\label{fig3}
\end{figure}

\subsubsection{The form of the collisional action}  \label{secphase} 
The collisional action $\Phi_{j}$ along the trajectory $j$ in \eq{S-m}
has some special properties which result from the form of the S-matrix
\eq{S-matrix}.  The asymptotically constant states are represented by
a constant action $J$ or quantum number $n$ and a constant momentum
$P$ for bound and free degrees of freedom respectively.  Hence, in the
asymptotic integrable situation with $\dot n = \dot P = 0$ before and
after the collision no action $\Phi_{j}$ is accumulated and the
collisional action has a well defined value irrespectively of the
actual propagation time in the asymptotic regions.  This is evident
from \eq{colacang} which is, however, not suitable for a numerical
realization of the collision.  The scattering process is much easier
followed in coordinate space, and more specifically
 for our collinear case, in radial
coordinates.  In the following, we will describe how to extract the
action according to \eq{colacang} from such a calculation in radial
coordinates (position $r$ and momentum $p$ for the target electron,
$R$ and $P$ for the projectile electron).  The discussion refers to
excitation processes to keep the notation simple but the result
\eq{acmom2} holds also for the other cases.  The collisional
action $\Phi$ can be expressed through the action in coordinate space
$\tilde\Phi$ by \cite{Mil70}
\be{acmom1}
\Phi(P,n;P',n') =  \tilde\Phi(P,r;P',r') + F_2(r',n') - F_2(r,n) ,
\ee
where  
\be{acort}
\tilde\Phi(P,r;P',r') = \lim_{t \to \, \infty \atop t' \to -\infty} \, \int 
\limits_{t'}^{t} \, d\tau \left[ -R \dot{P} + p \dot{r} \right] 
\ee
is the action in coordinate space and $F_2$ is the generator for 
the classical canonical transformation from the phase space
variables 
$(r,p)$ to  $(q,n)$ given by
\be{F2}
F_2(r,n) = {\rm{sgn}}(p) \, \int \limits_{r_i}^{r} 
\left(2 m \left[ \epsilon\left( n \right) - v \left(x \right) \right] 
\right)^{\frac{1}{2}} \, dx \, .
\ee
Here, $r_i$ denotes an inner turning point of an electron with energy $\epsilon(n)$ in the 
potential $v(x)$. Clearly, $F_2$ will not contribute if the
trajectory starts end ends at a turning point of the bound
electron. Partial integration of \eq{acort} transforms to momentum
space and yields a simple expression for the collisional action
in terms of spatial coordinates: 
\be{acmom2}
\Phi(P,n;P',n') = \lim_{t_i \to \, \infty \atop  
t_i' \to -\infty}  \, - \int 
\limits_{t_i'}^{t_i} \, d\tau 
\left[ R \dot{P} + r \dot{p} \right] \,.
\ee
Note, that $t_i'$ and $t_i$ refer to times where 
the bound electron is at an inner turning point and the generator $F_2$ vanishes. 
Phases  
determined according to \eq{acmom2} may still differ for the same path 
depending on its time of termination. However, 
the difference can only 
amount to integer multiples of the (quantized !) action 
\be{action} 
J = \oint p \, dr = 2 \pi \left(n + \frac{1}{2} \right) 
\ee 
of the bound electron with $\epsilon < 0$.
Multiples of $2 \pi$ for each revolution do not change the value of the S-matrix 
and the factor $ \frac{2\pi}{2}$ is compensated 
by the Maslov index.
In the case of an ionizing trajectory the action must be corrected 
for the logarithmic phase 
accumulated in Coulomb potentials \cite{Coul}. 

Summarizing this analysis, we fix the (in principle arbitrary) 
starting point of the trajectory to be an inner turning
point ($r_i'|p' = 0, \dot p' > 0$) which completes the initial
condition for the propagation of trajectories described in
section 2.2.1. In order to obtain the correct
collisional action \eq{colacang} in the form \eq{acmom2} 
we also terminate a trajectory at an inner turning point $r_i$
after the collision such that $\Phi$ is a continuous
function of the initial position $R'$. Although this 
is not necessary for the primitive 
semiclassical scattering probability which is only sensitive
to phase differences up to multiples of $J$ as mentioned above,
the absolute phase difference is needed for
 a uniformized  semiclassical
expression to be  discussed later. 

\subsection{Maslov indices}
 \label{maslnum}
 \subsubsection{Numerical procedure}
In position space the determination of the Maslov index is 
rather simple for an ordinary Hamiltonian with kinetic energy as
in \eq{hlanger}.
According to Morse's theorem the Maslov index is equal to the number of conjugate 
points along the trajectory. A conjugate point in coordinate space is defined by 
($f$ degrees of freedom, $(q_i,p_i)$ a pair of conjugate variables)
\be{focort}
det \left( M_{qp} \right) = det \left( \frac{\partial 
\left( q_1,\ldots,q_f \right)}{\partial 
\left(p_1',\ldots,p_f'\right)} \right) = 0.
\ee
The matrix $M_{qp}$ is the upper right part of the stability or 
monodromy matrix which is defined by 
\be{stabmatdef}
{\delta \vec{q}(t) \choose \delta \vec{p}(t)} \equiv M(t) \, 
{\delta \vec{q}(0) \choose \delta \vec{p}(0)} \, .
\ee
In general, the Maslov index $\nu_j$ in \eq{S-m} must be computed in the
same representation as the action. In our case this is the 
momentum representation in \eq{acmom2}.
However, the Maslov index in momentum space
is not simply the number of conjugate points in 
momentum space where $det \left(M_{pq} \right) = 0$. 
Morse's theorem relies on the fact that 
in position space the mass tensor 
$B_{ij} = \partial^2 H / \partial p_i \partial p_j$ 
is positive definite. This is not necessarily true for  
$D_{ij} = \partial^2 H / \partial q_i \partial q_j$ 
which is the equivalent of the mass tensor in momentum space.
How to obtain the correct Maslov index from 
the number of zeros of $\det \left(M_{pq} \right) = 0$ is described in 
\cite{Lev78}, a review about the Maslov index and its geometrical
interpretation is given in \cite{Lit92}.
\subsubsection{Phenomenological approach for two degrees of
freedom}
For two degrees of freedom, one can extract the
scattering probability directly from the deflection function 
without having to compute the stability matrix and its
determinant explicitly \cite{Ros94}. 
In view of this simplification it would
be desirable to determine the Maslov indices also directly
from the deflection function avoiding the complicated procedure
described in the previous section. This is indeed possible since
one needs only the correct {\it difference} of Maslov indices 
for a semiclassical scattering amplitude. 

A little thought reveals that trajectories starting from
branches in the deflection function of \fig{2} separated by an
extremum differ by one conjugate point. This implies that
their respective Maslov indices  differ by $\Delta\nu =1$. For this
reason it is convenient to divide the deflection function in
different branches, separated by an extremum. Trajectories of
one branch have the same Maslov index. Since there are two
extrema we need only two Maslov indices, $\nu_1 = 1$ and 
$\nu_2 = 2$. The relevance of just two values of Maslov indices
$(1,2)$ can be traced to the fact that almost all conjugate
points are trivial in the sense that they belong to 
turning points of bound two-body motion. 

We can assign the larger index $\nu_2 = 2$ to the 
trajectories which have passed one more conjugate point than
the others. 
As it is almost evident from their topology,
these are  the
trajectories with $d {\epsilon}/dR' > 0$ shown in 
 the upper row of \fig{1}. (They also have a larger
collisional action $\Phi_j$). The two non-trivial conjugate points
for these trajectories  compared to the single conjugate point  
 for orbits with initial conditions
corresponding to $d{\epsilon}/dR'<0$  can be understood looking at the
ionizing trajectories (b) and (e) of each branch in \fig{1}.
Trajectories from both branches have in common the turning point
for the projectile electron ($P = 0$). For trajectories 
of the lower row all other turning points belong to complete
two-body revolutions of a bound electron and may be regarded
as trivial conjugate points. However, for the trajectories from
the upper row there is one additional turning point (see, e.g.,
\fig{1}(b)) which
cannot be absorbed by a complete two-body revolution. It is
the source for the additional Maslov phase.

We finally remark that 
$d\epsilon/dR' > 0$ is equivalent to $dn/d\bar q<0$
of \cite{Mar72} leading to the same result as our considerations
illustrated above.

\section{Semiclassical scattering probabilities}
Taking into account the Pauli principle for the indistinguishable electrons 
leads to different excitation probabilities for singlet and triplet,
\bea
P^{+}_{\epsilon} (E) & = & \left| \, S_{\epsilon,\epsilon'}(E) + S_{E-\epsilon,\epsilon'}(E) 
\, \right|^2  
\nonumber \\ 
P^{-}_{\epsilon} (E) & = & \left| \, S_{\epsilon,\epsilon'}(E) - S_{E-\epsilon,\epsilon'}(E) \label{semprob} 
\, \right|^2 \, ,
\end{eqnarray}
where the probabilities are symmetrized a posteriori (see \cite{Joa75}). 
Here, $S_{\epsilon,\epsilon'}$ denotes the S-matrix for the excitation  branch,
calculated according to \eq{S-m}, while $S_{E-\epsilon,\epsilon'}$ represents 
the exchange  processes, at a fixed
energy $\epsilon < 0$, respectively.

Ionization probabilities are obtained by integrating the differential 
probabilities over the relevant energy range which is due to the symmetrization
\eq{semprob} reduced to $E/2$:
\be{ionint}
P^{\pm}_{ion} (E) = \int \limits_{0}^{E/2} P^{\pm}_{\epsilon} (E) \, d \epsilon \, .
\ee
\subsection{Ionization and excitation for singlet and triplet symmetry}
We begin with the ionization probabilities since they illustrate most clearly
the effect of the vanishing exchange branch for higher energies as illustrated in
\fig{3}. The semiclassical result for the Langer Hamiltonian \eq{hlanger} 
shows the effect of the vanishing exchange branch in the deflection function
\fig{3} which leads to merging $P^{\pm}$ probabilities at a finite energy $E$,
in clear discrepancy to the quantum result, see
\fig{4}. Moreover, the extrema in the deflection function lead to the sharp structures
below $E = 1$ a.u.. The same is true for the excitation probabilities
where  a discontinuity appears below $E = 1$ a.u. (\fig{5}). Note also 
that due to the violated unitarity in the semiclassical approximation 
probabilities can become larger than unity, as it is the case for the 
$n = 1$ channel.
\begin{figure}\hfill
\psfig{file=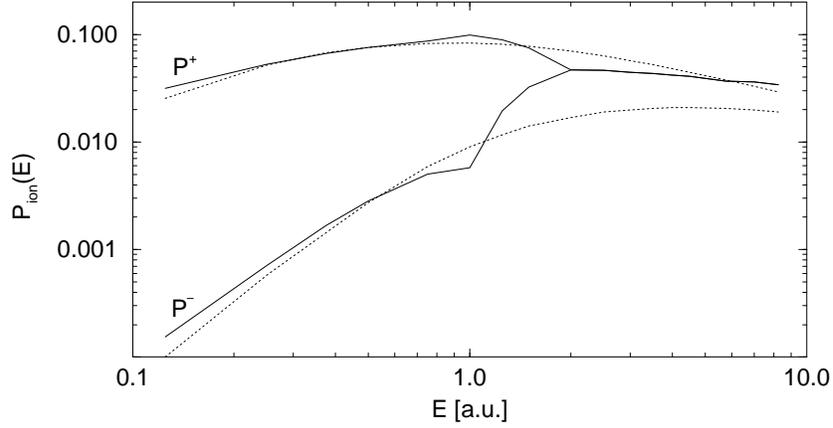,width=11cm}
\caption[]{Ionization probabilities for singlet and triplet according to \eq{ionint} 
with the Hamiltonian \eq{hlanger}
(solid line)  compared to quantum mechanical calculations (dotted line).} 
\label{fig4}
\end{figure}

\begin{figure}\hfill
\psfig{file=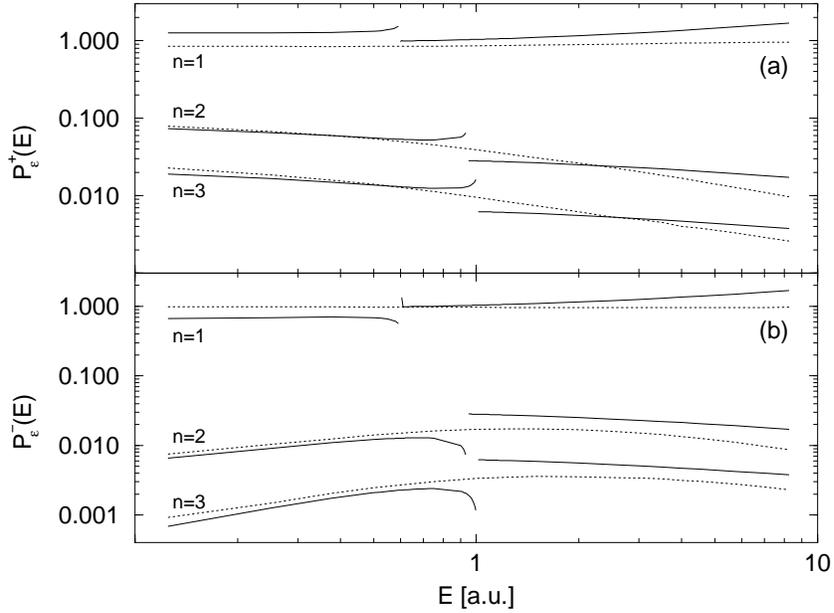,width=11.0cm}
\caption[]{Semiclassical excitation probabilities ($n=1,2,3$) according to \eq{semprob}
for singlet (part a) and triplet (part b) in the LSA (solid line) 
compared to quantum mechanical calculations (dotted line).} 
\label{fig5}
\end{figure}

Singlet and triplet excitation probabilities represent the most differential
scattering information for the present collisional system. Hence, the strongest
deviations of the semiclassical results from the quantum values can be expected.
Most experiments do not resolve the spin states and measure a spin-averaged 
signal. In our model this can be simulated by averaging the singlet and triplet
probabilities to
\be{probav}
P_{\epsilon} (E)  = \frac 12 (P^+_{\epsilon} (E)+ P^-_{\epsilon} (E)).
\ee
The averaged semiclassical probabilities may also be compared to the classical 
result which is simply given by
\be{probclas}
P^{CL}_{\epsilon}(E) = \sum_j({\cal P}^{(j)}_{\epsilon,\epsilon'}(E)+
{\cal P}^{(j)}_{\epsilon,E-\epsilon'}(E))
\ee
with ${\cal P}^{(j)}_{\epsilon,\epsilon'}(E)$ from \eq{probcl}.

\begin{figure}\hfill
\psfig{file=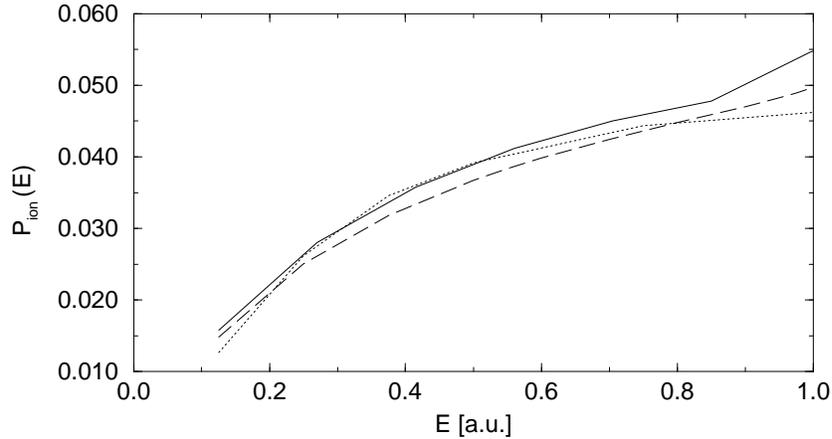,width=11.0cm}
\caption[]{Spin averaged quantum results for ionization (dotted line) compared
to averaged semiclassical probabilities (solid line) from \eq{probav} and 
classical probabilities (dashed line) from \eq{probclas}.}
\label{fig6}
\end{figure}

Figure \ref{fig6} shows averaged ionization probabilities. They are very similar to each
other, and indeed, the classical result is 
not much worse than the semiclassical result.

In \fig{7} we present the averaged excitation probabilities.
 Again, on can see the discontinuity resulting from the
extrema in the deflection function. As for ionization,
the spin
averaged semiclassical probabilities (\fig{7}b) are rather similar to the classical 
ones (\fig{7}a), in particular
the discontinuity is of the same magnitude as in the classical case and 
considerably more localized in energy than in the non-averaged quantities of \fig{5}. 

\begin{figure}\hfill
\psfig{file=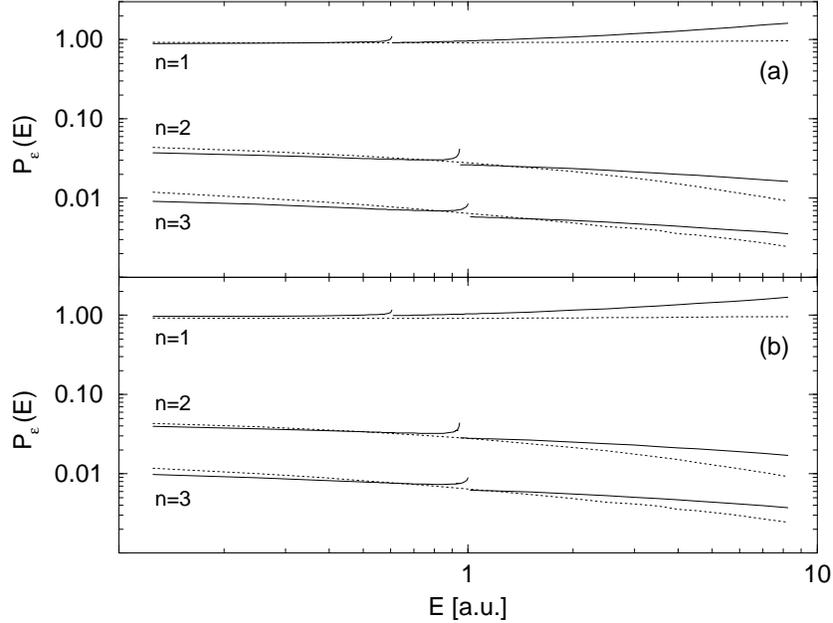,width=11.0cm}
\caption[]{Spin averaged quantum results (dotted line) for excitation ($n = 1,2,3$) compared
to classical probabilities (solid line, part a) from \eq{probclas} and
averaged semiclassical probabilities (solid line, part b) from \eq{probav}.}
\label{fig7}
\end{figure}

Clearly, the discontinuities are an artefact of the semiclassical approximation.
More precisely, they are a result of the finite depth of the two-body potential
in the Langer corrected Hamiltonian \eq{hlanger}. Around the extrema of the
deflection function the condition of isolated stationary points,
necessary to apply the stationary phase approximation which leads to 
\eq{S-m}, is not fulfilled. Rather, one has to formulate a uniform
approximation which can handle the coalescence of two stationary phase points.

\section{Uniformized scattering probabilities}
\label{usla} 
We follow an approach by Chester {\it et.\ al.} \cite{Che57}. The explicit expression
for the uniform S-matrix goes back to  Connor and Marcus \cite{Con71} who obtained for 
two coalescing trajectories $1$ and $2$
\be{uni}
S_{n,n'}(E) = {\rm Bi}^+ \left(-z \right) \, \sqrt{{\cal P}_{n,n'}^{(1)}(E)} \, 
e^{i\Phi_{1}+i\frac{\pi}{4}} \, + \, {\rm Bi}^- \left(-z \right) \, \sqrt{{\cal P}_{n,n'}^{(2)}(E)} 
\, e^{i\Phi_{2}-i\frac{\pi}{4}}  
\ee
where
\be{con}
{\rm Bi}^{\pm} \left(-z \right) = \sqrt{\pi} \left[z^{\frac{1}{4}} {\rm Ai} \left(-z \right) \mp
i z^{-\frac{1}{4}} {\rm Ai}' \left(-z \right) \right] \, 
e^{\pm i \left(\frac{2}{3} z^{\frac{3}{2}} - \frac{\pi}{4} \right)}
\ee
The argument $z = \left[ \frac{3}{4} \left(\Phi_2 - \Phi_1 \right) 
\right]^{\frac{2}{3}}$ of the Airy function
{\rm Ai(z)} contains the absolute phase difference. We assume that $\Phi_2 > \Phi_1$ which
implies for the difference of the Maslov indices that $\nu_2 - \nu_1 = 1$ 
(compare \eq{S-m} with \eq{uni} and \eq{largephi}).
Since the absolute phase difference enters \eq{uni} it 
is important to ensure that the action is a continuous function of $R'$ 
avoiding jumps of multiples of $2 \pi$, as already mentioned in section 2.2.2.
 For large phase differences \eq{S-m} is recovered since
\be{largephi}
\lim_{z \to \infty} {\rm Bi}^{\pm} \left(-z \right) = 1 \, .
\ee

Our uniformized S-matrix has been calculated by applying \eq{uni} to the two branches for exchange 
and excitation separately and adding or subtracting the results according to a singlet or triplet 
probability. In the corresponding probabilities  of \fig{8} the discontinuities of the non-uniform
results are indeed smoothed in comparison with \fig{5}. However, the overall agreement with the quantum probabilities is
worse than in the non-uniform approximation.
A possible explanation 
could lie in the construction of the uniform approximation. It works with an integral 
representation of the S-matrix, where the oscillating phase (the action) is mapped onto a cubic 
polynomial. As a result, the uniformization works best, if the deflection function can be 
described as a quadratic function around the extremum. Looking at \fig{2} one sees that 
this is true only in a very small neighborhood of the extrema because the deflection function is 
strongly asymmetric around these points.
We also applied a uniform approximation  derived by Miller \cite{Mil70a} which gave almost 
identical results. 

\begin{figure}\hfill
\psfig{file=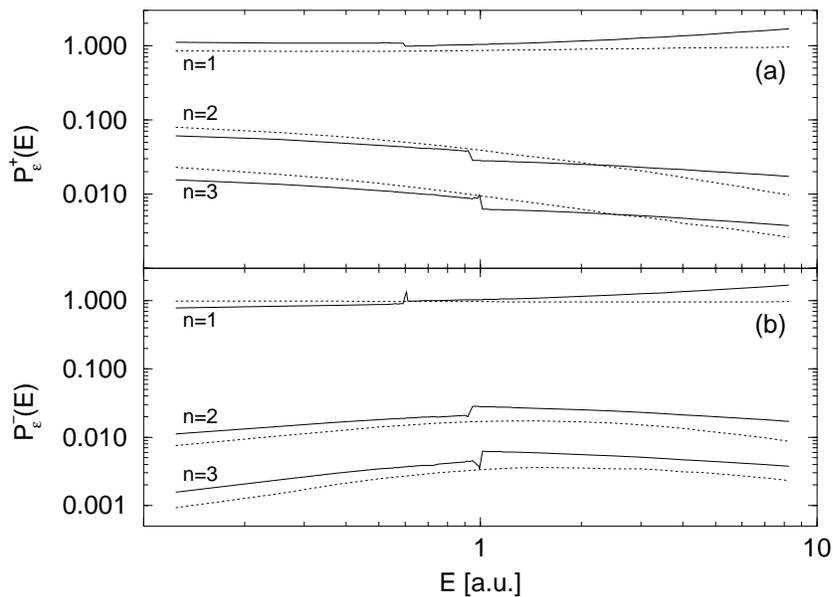,width=11cm}
\caption[]{Uniformized semiclassical excitation probabilities ($n=1,2,3$) according to \eq{uni} 
(solid line) for singlet (part a) and triplet (part b) 
compared to quantum mechanical calculations (dotted line).} 
\label{fig8}
\end{figure}

Finally, for the sake of completeness, the spin averaged uniform probabilities are shown in \fig{9}.
As can be seen, the discontinuities have  vanished almost completely. However, 
the general agreement with quantum mechanics is worse than for the standard  semiclassical 
calculations, similarly as for the symmetrized probabilities.

\begin{figure}\hfill
\psfig{file=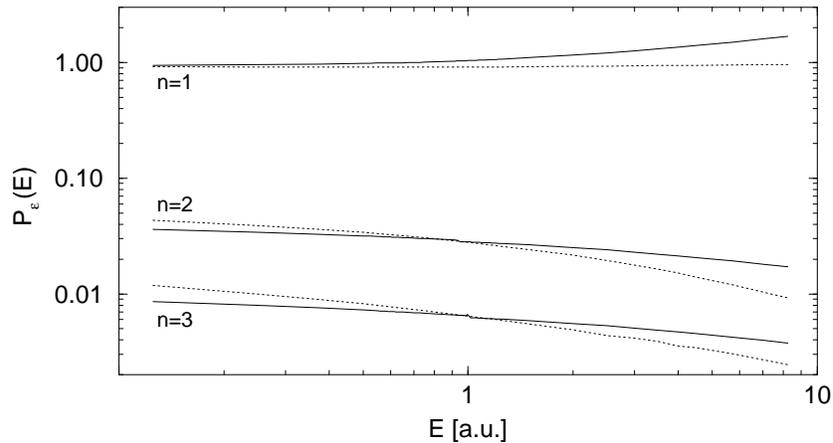,width=11cm}
\caption[]{Spin averaged uniformized excitation probabilities ($n=1,2,3$, solid line) compared
to quantum results (dotted line).}
\label{fig9}
\end{figure}

\section{Conclusion}
In this paper we have described  inelastic Coulomb scattering with a semiclassical
S-matrix. To handle the problem for this explorative study we have restricted
the phase space to the collinear arrangement of the two electrons reducing 
the degrees of freedom to one radial coordinate for each electron.
In appreciation of the spherical geometry we have applied the so called 
Langer correction to obtain the correct angular momentum quantization.
Thereby, a lower bound to the two-body potential is introduced which generates
a generic situation for bound state dynamics since the (singular) 
Coulomb potential is replaced by a potential bounded from below.
The finite depth of the two-body potential leads to singularities in the semiclassical
scattering matrix (rainbow effect) which call for a uniformization.

Hence, we have carried out and compared among each other
classical (where applicable), semiclassical, and uniformized semiclassical calculations 
 for the singlet, triplet and spin-averaged ionization and excitation 
probabilities. Two general trends may be summarized: Firstly, the simple semiclassical
probabilities are overall in better agreement with the quantum results for the
singlet/triplet observables than the uniformized results. The latter are only superior
close to the singularities. Secondly, for the (experimentally most relevant) spin-averaged
probabilities the classical (non-symmetrizable) result is almost as good as the
semiclassical one  compared to the exact quantum probability. This holds for 
excitation as well as for ionization. Hence, we conclude from our explorative study that
a full semiclassical treatment for spin-averaged observables
is probably not worthwhile since it does not produce
better results than the much simpler classical approach. Clearly, this conclusion has
to be taken with some caution since we have only explored a collinear, low dimensional
 phase space. 
\ack
We would like to thank A. Isele for providing us with the quantum results 
for the collinear scattering reported here.
This work has been supported by the DFG within the Gerhard Hess-Programm.


\end{document}